\documentclass[sigconf]{acmart}

\setcopyright{acmlicensed}
\copyrightyear{2018}
\acmYear{2018}
\acmDOI{XXXXXXX.XXXXXXX}

\usepackage{caption}
\usepackage{subcaption}
\usepackage{multirow}
\usepackage{graphicx}
\usepackage{float}
\usepackage{todonotes}
\copyrightyear{2026}
\acmYear{2026}
\setcopyright{cc}
\setcctype{by-nc-nd}
\acmConference[AHs 2026]{The Augmented Humans International Conference 2026}{March 16--19, 2026}{Okinawa, Japan}
\acmBooktitle{The Augmented Humans International Conference 2026 (AHs 2026), March 16--19, 2026, Okinawa, Japan}
\acmPrice{}
\acmDOI{10.1145/3795011.3795059}
\acmISBN{979-8-4007-2351-3/2026/03}

\begin{document}

\title[Abstraction Beats Realism]{Abstraction Beats Realism: Physiological Visualizations Enhance Arousal Synchrony in VR Concert Recreations}

\author{Xiaru Meng}
\authornote{Both authors contributed equally to this research.}
\email{xiarumeng2019@outlook.com}
\orcid{0000-0003-1677-4839}
\author{Yulan Ju}
\authornotemark[1]
\email{yulan-ju@kmd.keio.ac.jp}
\orcid{0000-0002-5746-3933}
\affiliation{%
  \institution{Keio University Graduate School of Media Design}
  \city{Yokohama}
  \country{Japan}
  \postcode{223-8526}
}

\author{Matthias Hoppe}
\orcid{0000-0002-5098-4824}
\email{matthias.hoppe@kmd.keio.ac.jp}
\affiliation{%
  \institution{Keio University Graduate School of Media Design}
  \city{Yokohama}
  \country{Japan}
  \postcode{223-8526}
}

\author{Jiawen Han}
 \email{hanjiawen@kmd.keio.ac.jp}
 \orcid{0000-0002-0714-6227}
\affiliation{%
  \institution{Keio University Graduate School of Media Design}
  \city{Yokohama}
  \country{Japan}
  \postcode{223-8526}
}

\author{Yan He}
\orcid{0000-0003-2292-0863}
\email{yann@kmd.keio.ac.jp}
\affiliation{%
  \institution{Keio University Graduate School of Media Design}
  \city{Yokohama}
  \country{Japan}
  \postcode{223-8526}
}

\author{Kouta Minamizawa}
\orcid{0000-0002-6303-5791}
\email{kouta@kmd.keio.ac.jp}
\affiliation{%
  \institution{Keio University Graduate School of Media Design}
  \city{Yokohama}
  \country{Japan}
  \postcode{223-8526}
}

\author{Kai Kunze}
\email{kai@kmd.keio.ac.jp}
\orcid{0000-0003-2294-3774}
\affiliation{%
  \institution{Keio University Graduate School of Media Design}
  \city{Yokohama}
  \country{Japan}
  \postcode{223-8526}
}

\renewcommand{\shortauthors}{Meng and Ju, et al.}

\begin{abstract}
Live cultural experiences like concerts generate shared physiological arousal among audience members, a collective resonance that contributes to their emotional power. Recreating such experiences in virtual reality therefore requires not just audiovisual fidelity, but reproduction of this physiological dimension. Yet current VR evaluation methods rely on post-hoc self-reports that interrupt immersion and cannot capture moment-to-moment arousal dynamics. We propose cross-temporal physiological synchrony as an unobtrusive methodology for evaluating VR cultural recreations: measuring how closely a VR participant's arousal patterns align with those of the original live audience. In a two-phase study, we recorded electrodermal activity from 40 live concert attendees, then created three VR recreations with varying abstraction levels (realistic 360-degree video, mixed video-plus-visualization, and fully abstract physiological representations) and measured synchrony with 22 laboratory participants using Dynamic Time Warping. Contrary to assumptions favoring realism, abstract visualizations achieved the strongest synchrony with live audiences (Symbolic: r(258) = .96, p < .001; Mixed: r(258) = .92; Video: r(258) = .91), with substantial effect sizes (Cohen's d = 0.94 Symbolic vs. Video, d = 0.73 Symbolic vs. Mixed). During musical climaxes, the abstract condition maintained correlation while realistic video showed none. These findings suggest that abstract physiological representations may be more effective than realistic footage for evoking authentic collective engagement in VR cultural recreations.
\end{abstract}



\begin{teaserfigure}
  \includegraphics[width=\textwidth]{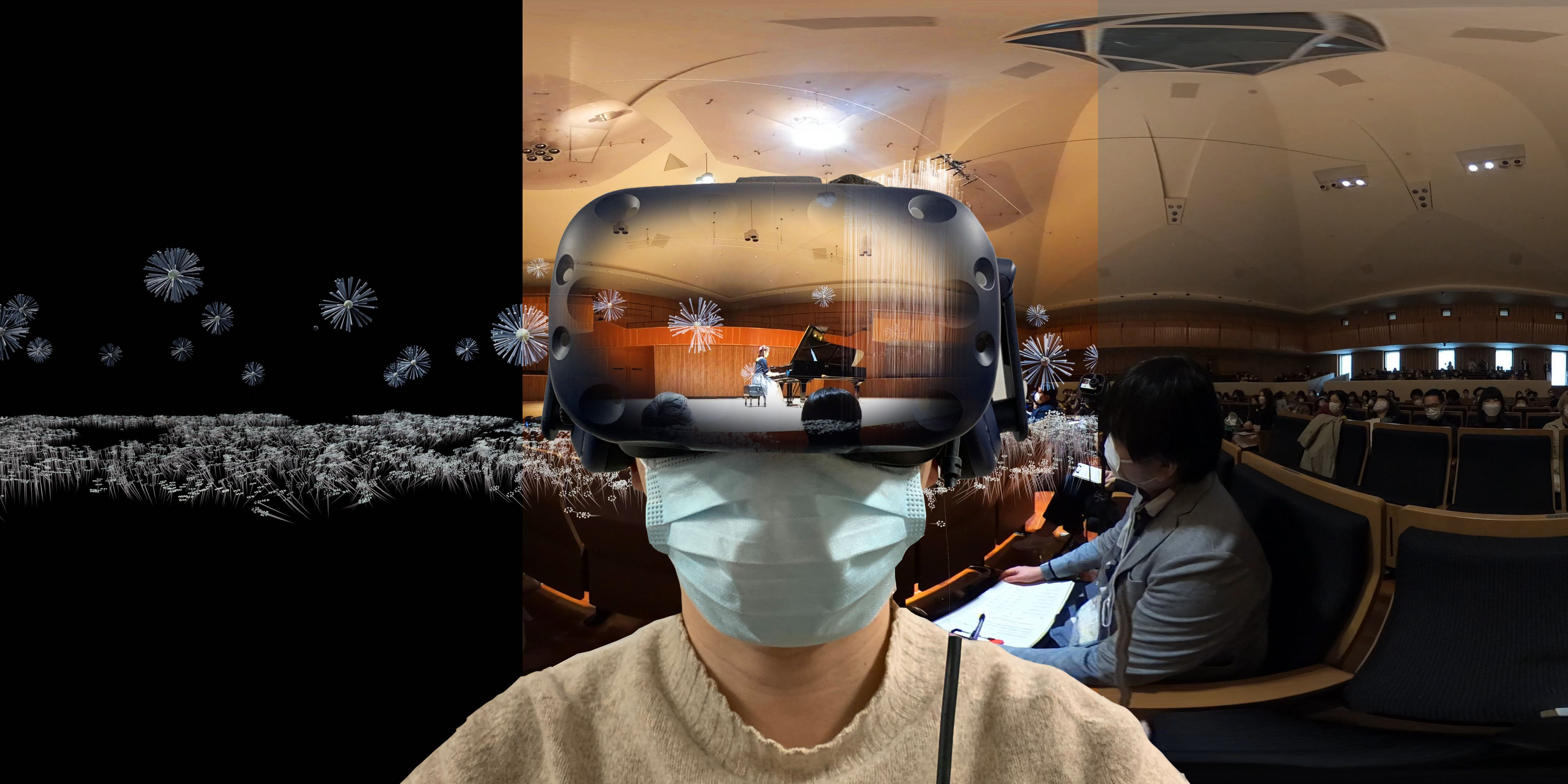}
  \caption{Cross-temporal physiological synchrony in VR concert recreation. We compare EDA patterns between live concert audiences (N=40) and VR participants (N=22) across three visualization conditions with varying abstraction levels.}
  \Description{}
  \label{fig:teaser}
\end{teaserfigure}

\maketitle

\section{Introduction}
Live performances such as theater and concerts are complex, social and physiological experiences. While such performances themselves can be recorded, capturing the atmosphere, involvement and physiological responses of their audiences proves to be a more challenging task. This raises methodological questions: how can we assess whether VR recreations achieve authentic engagement, and can the physiological patterns observed in live audiences serve as reference signals for evaluating such recreations?

The recreation of real-life experiences is often the focus of virtual reality research, games, and experiences. To evaluate user responses to these recreations, researchers use concepts such as presence~\cite{slater2003note, lee2004presence}, game experience~\cite{ijsselsteijn2013game}, and liveness~\cite{auslander2008liveness}. While researchers often measure these concepts via questionnaires deployed before and after events, they could also measure some responses via physiological devices~\cite{10.1145/3351247}.

Even for seated, pre‑recorded content, VR affords immersive, embodied viewing with spatial context, scale, and visual occlusion distinct from desktop video. We scope the present work to non‑interactive replay and explicitly examine how integrating prior audience physiology into the VR scene affects a viewer's arousal synchrony with the original audience.

The other attendees and the social atmosphere they create form an integral part of any live performance. When we are with people, we often observe interpersonal physiological synchrony (our body signals follow each other)~\cite{gordon2020relationship}, ranging from simple eye blinks and head nods in conversations to heart rate features in music performances and EDA in theater plays~\cite{shoda2016live,gupta2019blink,silveira2013predicting,he2022frisson}.

Cultural events require more than physical co-presence. They need deeper physiological information integrated into recreations. Music concerts provide excellent examples because they invoke measurable arousal responses~\cite{bernardi2006cardiovascular, bachrach2015audience} and connect people through shared physiological patterns~\cite{czepiel2021synchrony}. Different people exhibit similar autonomic nervous system reactions to musical stimuli~\cite{KHALFA2002145}.

Live concerts offer unique physiological and interpersonal experiences. VR recreations struggle to capture the same physiological response patterns. We compress the live event experience into 5-minute VR sessions and examine whether they show similar arousal patterns.

We propose cross-temporal physiological synchrony 
as an evaluation methodology for VR experiences. Current evaluation relies on questionnaires that disrupt experiments or suffer from retrospective bias~\cite{conner2012trends,doherty2018construal,chang2024perceived}.

\emergencystretch=1em
EDA reflects sympathetic arousal rather than specific emotions~\cite{braithwaite2013guide}, 
while excitement and anxiety produce comparable responses despite opposite valence. We measure cross-temporal physiological alignment as shared autonomic activation, not emotional contagion.

\emergencystretch=0em

\subsection*{Research Questions}
Given these considerations, we investigate three interconnected questions:

\begin{itemize}
    \item \textbf{RQ1 (Methodological Validity):} Can cross-temporal physiological synchrony, the alignment of VR participants' arousal patterns with those of a live audience, serve as a valid measure for evaluating cultural recreations in VR? We examine whether this approach yields consistent, interpretable signals that complement traditional self-report measures.

    \item \textbf{RQ2 (Abstraction Effects):} How does the level of visual abstraction in representing audience physiological data affect cross-temporal arousal synchrony? We compare realistic video, mixed representations, and fully abstract symbolic visualizations to understand whether realism is necessary, or even beneficial, for achieving physiological alignment.

    \item \textbf{RQ3 (Subjective-Physiological Relationship):} What is the relationship between physiological synchrony and participants' subjective experience of presence, arousal, and connection? This addresses whether physiological measures capture something distinct from what questionnaires assess, or whether they simply mirror self-reported experience.
\end{itemize}

\begin{figure*}[t]
  \centering
  \includegraphics[width=\linewidth]{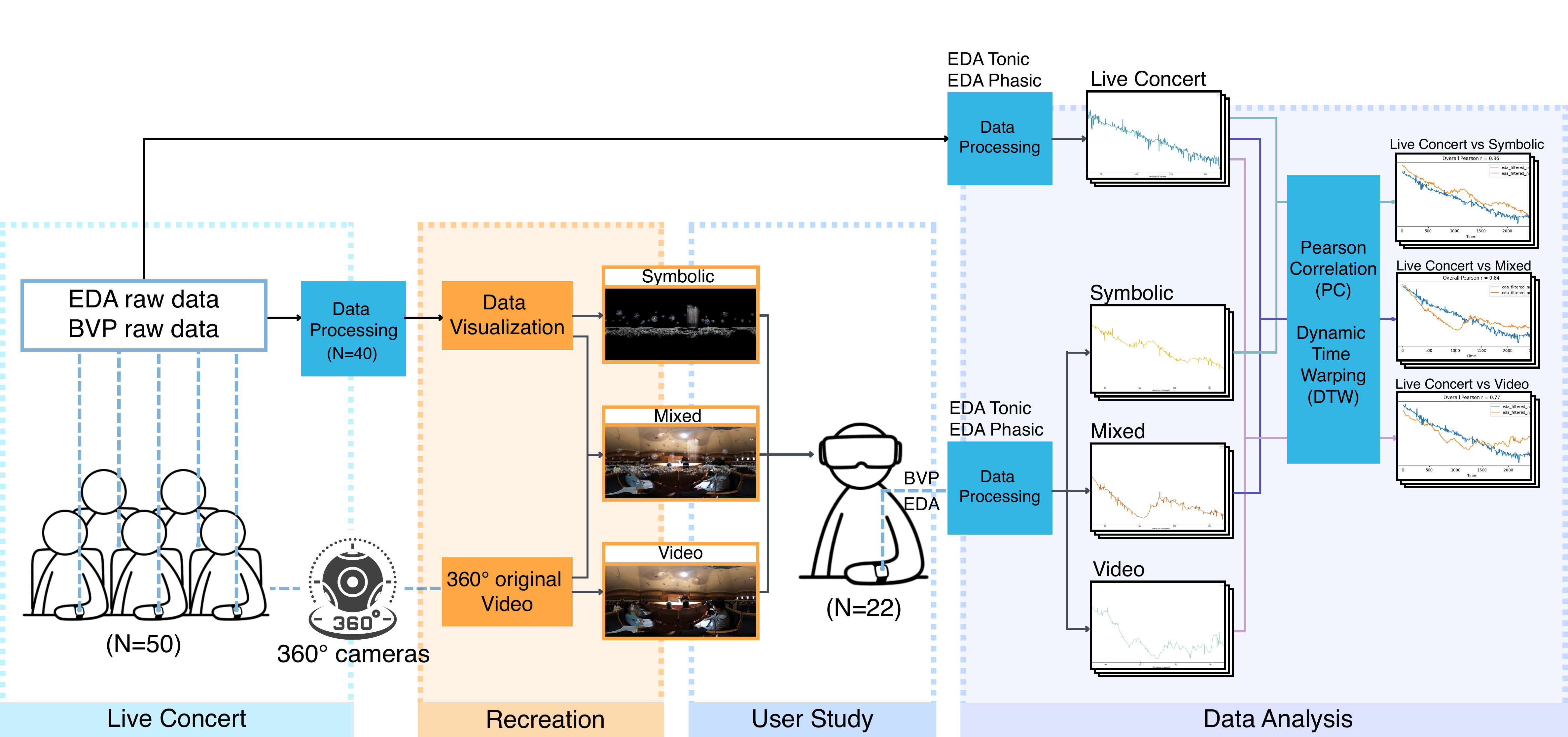}
  \caption{The pipeline processes data collected during both a live concert and a VR experiment.
  During the live concert we recorded a 360° video and physiological data of the audience. In the user study the participant is presented with three types of visualizations, the \textsc{Video} recording, \textsc{Symbolic} condition visualizing physiological data of the concert audience and \textsc{Mixed}, a combination of these two. The physiological data of the VR participant is recorded, analyzed and compared.}
  \label{workflow}
  \Description{}
\end{figure*}

We prototype a system that encodes physiological data from 40 concert audience members into VR visualizations (Figure~\ref{workflow}), investigating whether abstract representations produce cross-temporal physiological alignment. We make three contributions:

\smallskip

\textbf{1) Methodological Innovation:} 
We introduce a cross-temporal evaluation methodology that uses physiological synchrony to assess arousal alignment in VR cultural recreations. We employ Dynamic Time Warping analysis to measure alignment between live concert audience responses and VR participants. This provides an alternative to traditional questionnaire-based evaluations. Our approach captures moment-to-moment autonomic responses without disrupting the experience. We demonstrated very strong correlations between live and VR audiences ($r(258) = .96$, $p < .001$ for the most effective condition).

\smallskip

\textbf{2) Technical System and Empirical Validation:} 
We developed a modular 3D visualization system that transforms physiological data from 40 live concert audience members into ambient biometric interfaces for VR. Our within-subjects study ($N = 22$) across three abstraction levels demonstrates that abstract physiological visualizations showed higher physiological synchrony compared to realistic recreations. The \textsc{Symbolic} condition showed significantly higher synchrony than \textsc{Video} (Cohen's $d = 0.94$, $p < .001$) and \textsc{Mixed} conditions (Cohen's $d = 0.73$, $p = .021$). Peak synchrony occurred during musical climax periods (100--160s) with medium effect sizes across all conditions ($\eta^2_p = .069$--$.079$).

\smallskip

\textbf{3) Design Insights and Implications:} 
Abstraction is associated with higher perceived social connection and arousal synchrony. Removing visual realism while preserving physiological patterns corresponded to higher autonomic alignment. The abstract condition achieved the highest tonic EDA correlation ($r(258) = .96$) compared to \textsc{Mixed} ($r(258) = .92$) and \textsc{Video} ($r(258) = .91$) conditions. Contrary to expectations, higher abstraction correlates with stronger arousal synchrony. This provides initial design insights for biometric-responsive environments and contributes a multi-modal dataset of synchronized physiological and environmental data.

\section{Related Work}

Cultural experiences create moments of collective engagement through shared presence and physiological alignment. However, temporal and spatial constraints limit access to these experiences. We ask whether virtual reality can help recreate aspects of these moments by integrating prior audience physiology. Our cross-temporal approach captures live audience physiological responses and integrates them into VR replays, and we evaluate whether abstract representations of collective arousal can elicit similar synchrony patterns in subsequent VR audiences.

\subsection{VR Cultural Recreation and Presence}

Virtual reality's capacity to recreate cultural experiences has been studied in relation to presence theory. Witmer and Singer's presence framework~\cite{witmer1998measuring} identifies perceptual immersion, attentional focus, and involvement as factors relevant to culturally significant events. Slater's complementary work~\cite{slater2003note, slater2009place} demonstrates that convincing VR depends on both technological fidelity and psychological engagement, with place illusion and plausibility illusion as core elements.

Cultural VR applications span diverse domains. Educational recreations foster empathetic understanding through embodied experience~\cite{ahn2016experiencing}, while virtual museums help to visualize and spatialize abstract concepts~\cite{su12072698}. However, these studies predominantly focus on visual and auditory factors. Outside VR, audience physiology has predicted responses to film content~\cite{silveira2013predicting} and shown entrainment during live performance~\cite{bachrach2015audience}, and collective EDA acquisition in the wild is feasible with wearables~\cite{bota2020wearable}.

Concert recreation represents a particularly challenging domain. Music exerts profound physiological effects, triggering autonomic responses including changes in heart rate, skin conductance, and respiration that reflect that of emotional arousal~\cite{scherer2005emotions, zhang2018pmemo}. Research on virtual concerts reveals mixed outcomes: while users report it to be ‘the future of the music industry’~\cite{onderdijk2023concert}, VR musical experiences lack the embodied engagement of live performance~\cite{venkatesan2023feeling}. This gap between reported presence and physiological authenticity motivates investigation of physiologically-informed VR design.

Recent work has explored incorporating physiological data into VR concerts to bridge this gap. Horie et al.~\cite{horie2017participation} demonstrated VR concert systems that generate visual effects from brainwaves and heartbeat, finding that users felt correspondence between their excitement and the visualizations. Research on frisson detection demonstrates that collective aesthetic responses can be captured in musical contexts~\cite{he2021frisson, he2022frisson}, while physiological sonification systems show promise for enhancing audience-performer connections~\cite{sugawa2021boiling}. Preliminary work by Meng et al.~\cite{meng2023towards} explored visualizing audience physiological data in VR concert replays, investigating whether such visualizations could enhance emotional response. However, our recent investigations reveal placebo effects where users' knowledge that data represents ``real'' audience responses impacts engagement, even when fabricated~\cite{10.1145/3706598.3713594}. Underscoring the importance of validating whether authentic physiological integration provides measurable benefits.

Presence research increasingly recognizes social dimensions, with co-presence theory highlighting how shared virtual experiences create different engagement patterns than solitary ones~\cite{zhao2003social}. Collaborative VR experiences validate this extension, with shared tasks showing measurable increases in social presence and collaborator awareness~\cite{10494151}. Yet little work examines whether virtual recreations can preserve collective physiological patterns that define authentic cultural experiences.

\subsection{Physiological Measurement in HCI and VR}

Physiological sensing has become instrumental in HCI for understanding user states that traditional interaction measures cannot capture. Wearable biosensors quantify autonomous nervous system responses during interactive experiences~\cite{swindells2006case, garbarino2014empatica}, providing objective measures of engagement, stress, and emotional arousal~\cite{andreassi2010psychophysiology}. This capability is particularly valuable in VR contexts where traditional usability metrics may not adequately capture embodied experiences.

Electrodermal activity (EDA) and blood volume pulse (BVP) serve as reliable indicators of sympathetic nervous system activation. EDA comprises phasic (rapid, stimulus-responsive) and tonic (slower baseline) components~\cite{boucsein2012electrodermal}, with both increasing during heightened emotional arousal. EDA's linearity has proven useful for emotion recognition~\cite{schmidt2018wearable, choi2011development}. Recent VR studies demonstrate that EDA reliably differentiates virtual task complexity~\cite{chiossi2022virtual} and predict cognitive load~\cite{10937342}, while interactive VR tasks can reliably enhance positive emotions through EEG-based measurements~\cite{10494145}. BVP measurements capture cardiovascular responses reflecting both emotional valence and arousal intensity~\cite{levenson2003blood}.

Methodological challenges persist. Motion artifacts from head-mounted displays can compromise signal quality~\cite{tremmel2019eeg}, 
while suitable measures may be influenced by movement and signal robustness, complicating interpretation of these measures across individuals~\cite{ahmadi2023cognitive}.
Others report that challenges arise through significant inter-individual variability in physiological responses to workload~\cite{wei2025cognitive}.
More fundamentally, most VR physiological studies focus on individual responses rather than collective patterns, limiting understanding of how VR might recreate social physiological dynamics.

The integration of physiological feedback into VR represents an emerging frontier. 
Research on adaptive visulisations demonstrates that real-time physiological feedback can improve the user experience~\cite{chiossi2022virtual}, 
while H\"{o}\"{o}k's ``Somaesthetic Appreciation'' framework~\cite{hook2016somaesthetic} suggests that physiological rhythms, "following the rhythm of the body", can serve as meaningful interaction modalities. EDA has been employed to assess engagement with artistic performance videos~\cite{latulipe2011love} and to gage audience reactions to varied content~\cite{silveira2013predicting}. However, these approaches typically focus on individual biofeedback rather than collective physiological patterns.

\subsection{Physiological Synchrony and Cross-temporal Analysis}

Physiological synchrony indicates social connection and shared experience quality~\cite{cheong2020synchronized}. Levenson and Gottman~\cite{levenson1985physiological} the established the use of physiological and affective measurements in research on social interaction; 
subsequent work extended this to therapeutic~\cite{marci2007physiologic}, educational~\cite{dikker2017brain}, and performance contexts~\cite{konvalinka2011synchronized, bachrach2015audience}. Researchers employ correlation approaches~\cite{hernandez2014using} and dynamic time warping~\cite{giorgino2009computing} to quantify synchrony across contexts including VR~\cite{moharana2023physiological}.

Critical gaps remain: most work examines contemporaneous synchrony between co-present individuals, while cross-temporal alignment between groups remains unexplored. 
Group synchrony may differ from dyadic mechanisms~\cite{konvalinka2011synchronized}, which is an important consideration for collective VR experiences.

\subsection{Evaluation Methodologies for Immersive Cultural Experiences}

Traditional evaluation methods suffer from recall bias~\cite{conner2012trends}, attention diversion~\cite{larson2014experience, toet2019emojigrid}, 
or the need to interrupt the VR experience by either removing the headset between conditions or to fill out in-VR questionnaires~\cite{schwind2019using}. 
Timeline-corresponding retrospective approaches~\cite{potts2025retrosketch} still depend on conscious recall. These limitations are problematic for cultural experiences where social and emotional dimensions may be difficult to articulate.

\subsection{Research Gaps}

Presence theory~\cite{witmer1998measuring, slater2003note} lacks development around collective presence and temporal preservation. VR physiological research emphasizes individual responses~\cite{10494076, 10937342} rather than collective patterns. Scherer's Component Process Model~\cite{scherer2005emotions, scherer2009dynamic} suggests similar event appraisals produce synchronized responses, but cross-temporal VR applications remain unexplored.

Our work introduces cross-temporal physiological synchrony as a framework for evaluating VR cultural recreation, extending presence theory to include collective physiological dimensions.

\section{Recreation: Bringing a Live Concert to VR}
We developed a system that transforms live concert physiological data into VR visualizations, enabling study of whether abstract representations enhance cross-temporal physiological alignment. Our prototype consists of three phases: field data collection (N=50, 40 retained after cleaning), a modular 3D visualization system integrating audience data, and three VR conditions with increasing abstraction levels.

\vspace{-0.2cm}

\subsection{Live concert}
\label{sec:concert}

\subsubsection{Data collection}We recorded physiological data from 50 audience members (25 female, 20 male, 5 unspecified; ages 7--73, $M = 33.48$, $SD = 14.65$) seated near the stage. A GoPro OMNI 360 rig captured 360-degree video. Shimmer3 GSR+ sensors~\cite{burns2010shimmer} recorded EDA and BVP data, following established finger-placement protocols~\cite{van2012emotional}. IRB approval and informed consent were obtained.

\subsubsection{Performance}
We recorded a 260-second solo piano performance of Beethoven's Sonata No. 28 (first movement), performed by a professional pianist with 26 years of experience. The piece progresses from a tranquil opening through a climactic section (100--160s) before returning to a peaceful state. The performer identified the 100--160s segment as the emotional climax, which informed our time-segmented physiological analysis.

\subsection{Recreation process}

This section presents a prototype for visualizing physiological data collected from live concert audiences. Our goal is to trigger physiological synchrony in VR users by integrating concert audience physiological data into the recreation to convey arousal patterns.

\subsubsection{Data processing}
We recorded physiological signals from 50 audience members who attended the concert. After data cleaning procedures, we retained 40 complete EDA and BVP datasets for visualization.
Data reduction occurred for two reasons: First, some wearable devices did not record BVP or EDA data, so we selected only complete datasets to present both signal types for each audience member. Second, after integrating the data with the initial model and conducting an inspection, we discarded all EDA data that did not conform to Boucsein's definition of EDA responses~\cite{boucsein2012electrodermal,hernandez2014using}. We applied filtering processes to both raw EDA and BVP data~\cite{virtanen2020scipy}. We normalized the EDA and BVP data to fit the dimensions of the paired model and resampled them at 30 Hz to align with the animation setting (FPS=30).

\subsubsection{Design Process}
We developed the visualization through iterative design informed by feedback from HCI researchers, musicians, and designers at a public exhibition. Key design principles included: (1) visual effects must be clear and easily discernible; (2) effects should reflect the performance's aesthetic; and (3) visualizations should highlight physiological characteristics.

\begin{figure}[t]
  \centering
  \includegraphics[width=\linewidth]{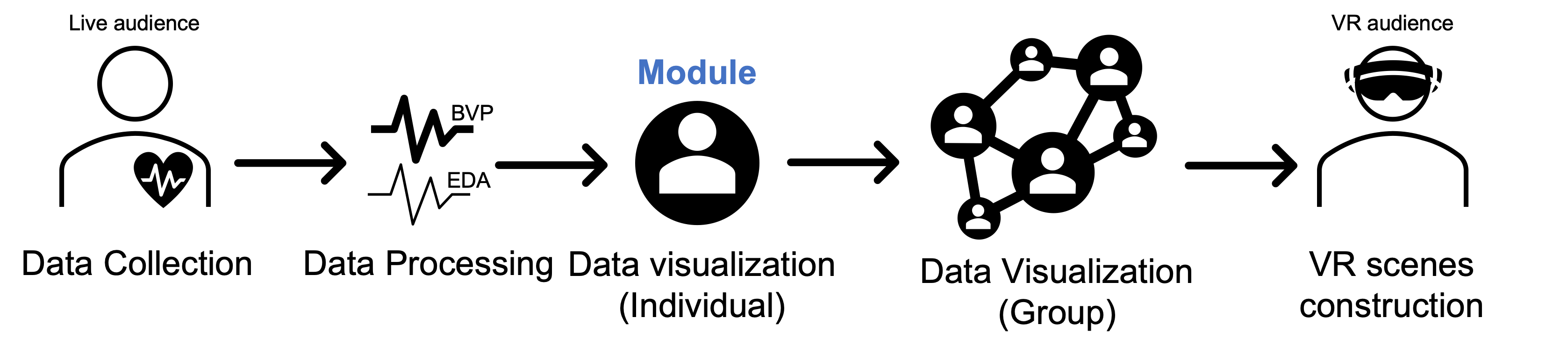}
  \caption{The visualization pipeline processes recorded physiological data of audience members into visualizations displayed in multiple scenes for the VR user.}
  \label{figframe}
  \Description{}
\end{figure}

Our design draws on modularity and bio-art concepts (see Figure~\ref{figframe}). We transformed the 40 complete BVP and EDA datasets into individual 3D models, each representing one audience member's physiological state. Organic forms (dandelion spheres, grass) were chosen because perceiving expressive biological cues like heartbeats enhances physiological insight and interpersonal connection~\cite{winters2021can}.

The visualization system was developed in Blender using the Cycles Render Engine. Organic nature metaphors were selected based on biophilic design research showing that natural fractal patterns reduce stress~\cite{taylor2006fractal} and that intuitive physiological visualizations effectively communicate bodily states~\cite{norooz2015bodyvis, neupane2024momentary}.

\begin{figure*}[t]
  \setlength{\belowcaptionskip}{-0.3cm} 
  \setlength{\abovecaptionskip}{-0cm}
  \includegraphics[width=0.95\textwidth]{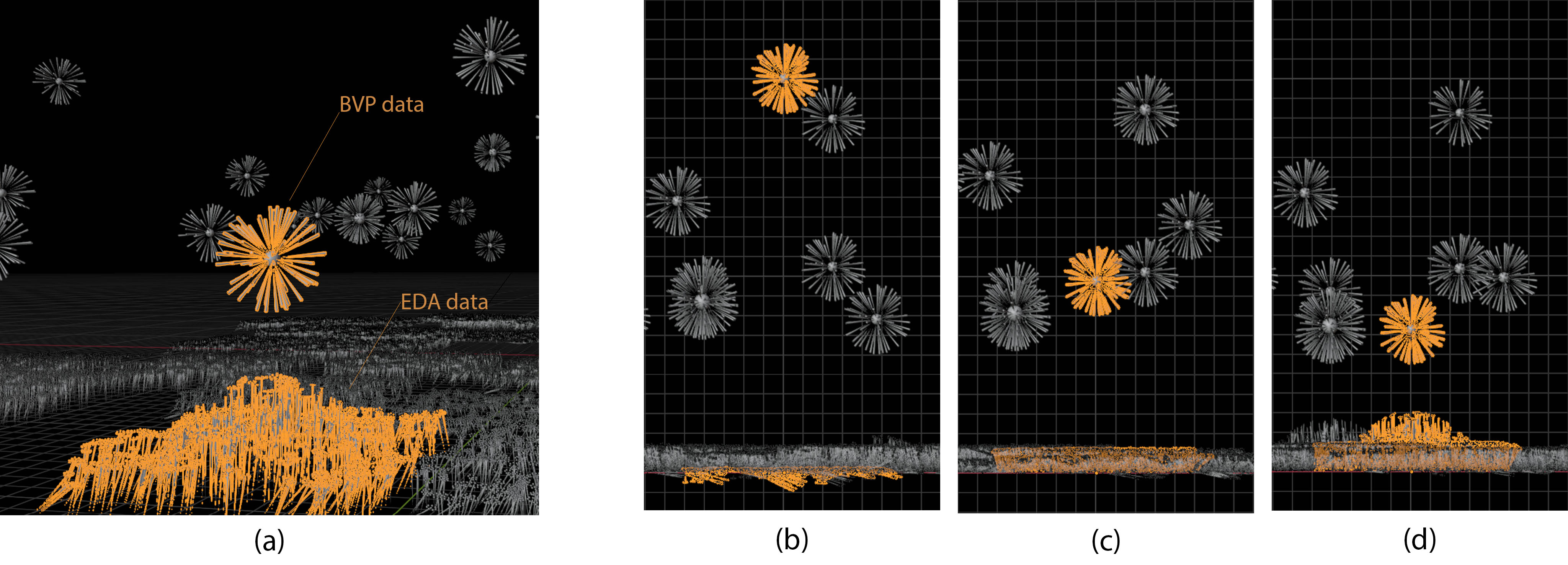}
  \caption{Each spherical model in the sky (flower) and batch of lines on the ground (grass) is a representation of a single audience member's response.
  The flower represents an audience member's heartbeat and changes in size accordingly (BVP).
The vertical position of the flower and height of the grass changes with the level of arousal (EDA).
The example visualizations show: (a) orange highlighted flower and grass visualize the BVP and EDA for one audience member; (b) for low EDA values; (c) for medium EDA values; (d) for high EDA values}
  \label{fig1}
  \Description{}
\end{figure*}

\subsection{Conditions in VR: Levels of visual abstraction}
\begin{figure}
     \centering
     \begin{subfigure}[b]{0.33\textwidth}
         \centering
         \includegraphics[width=\textwidth]{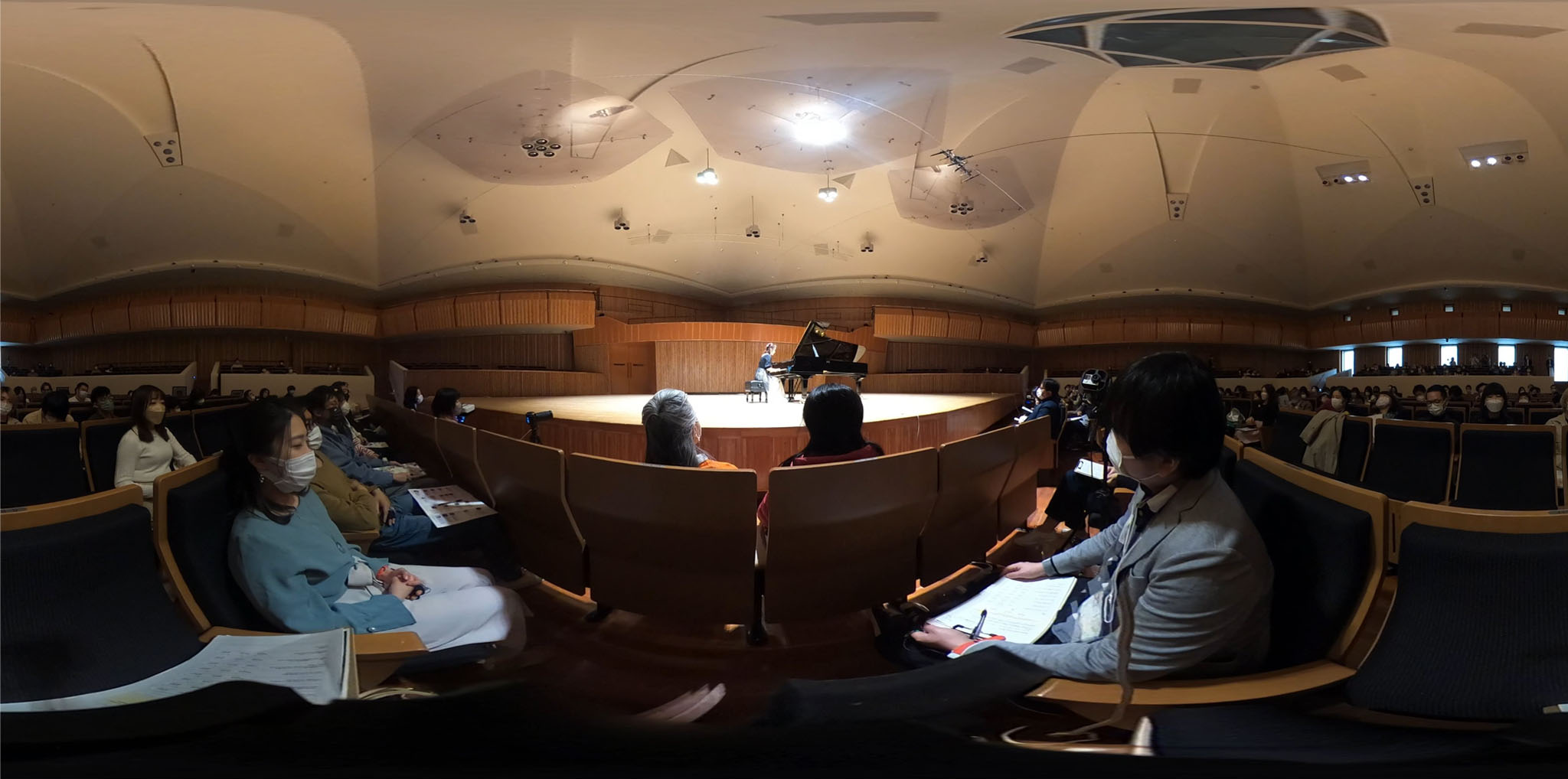}
         \caption{}
         \label{fig:y equals x}
     \end{subfigure}
     \hfill
     \begin{subfigure}[b]{0.33\textwidth}
         \centering
         \includegraphics[width=\textwidth]{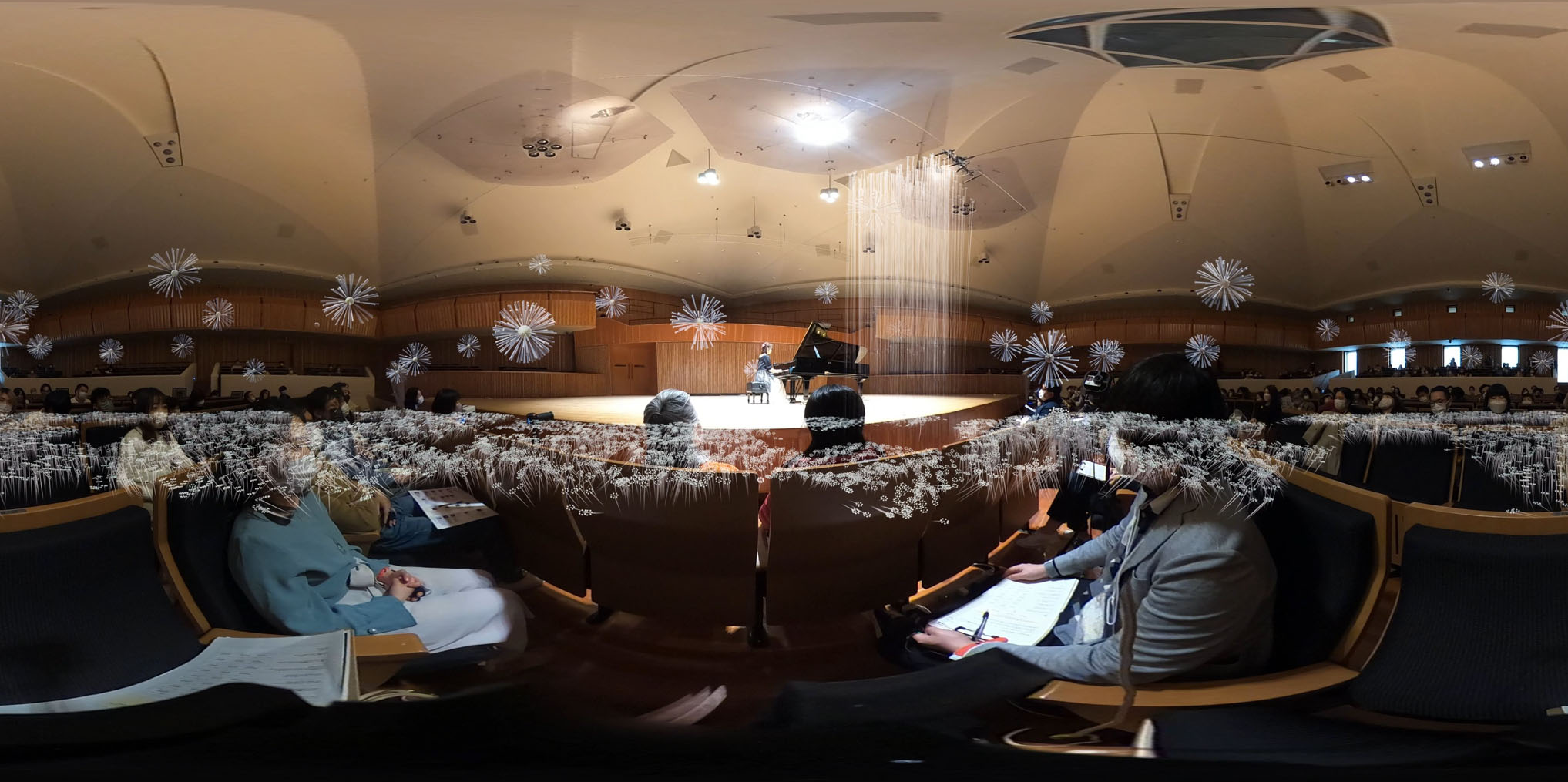}
         \caption{}
         \label{fig:three sin x}
     \end{subfigure}
     \hfill
     \begin{subfigure}[b]{0.33\textwidth}
         \centering
         \includegraphics[width=\textwidth]{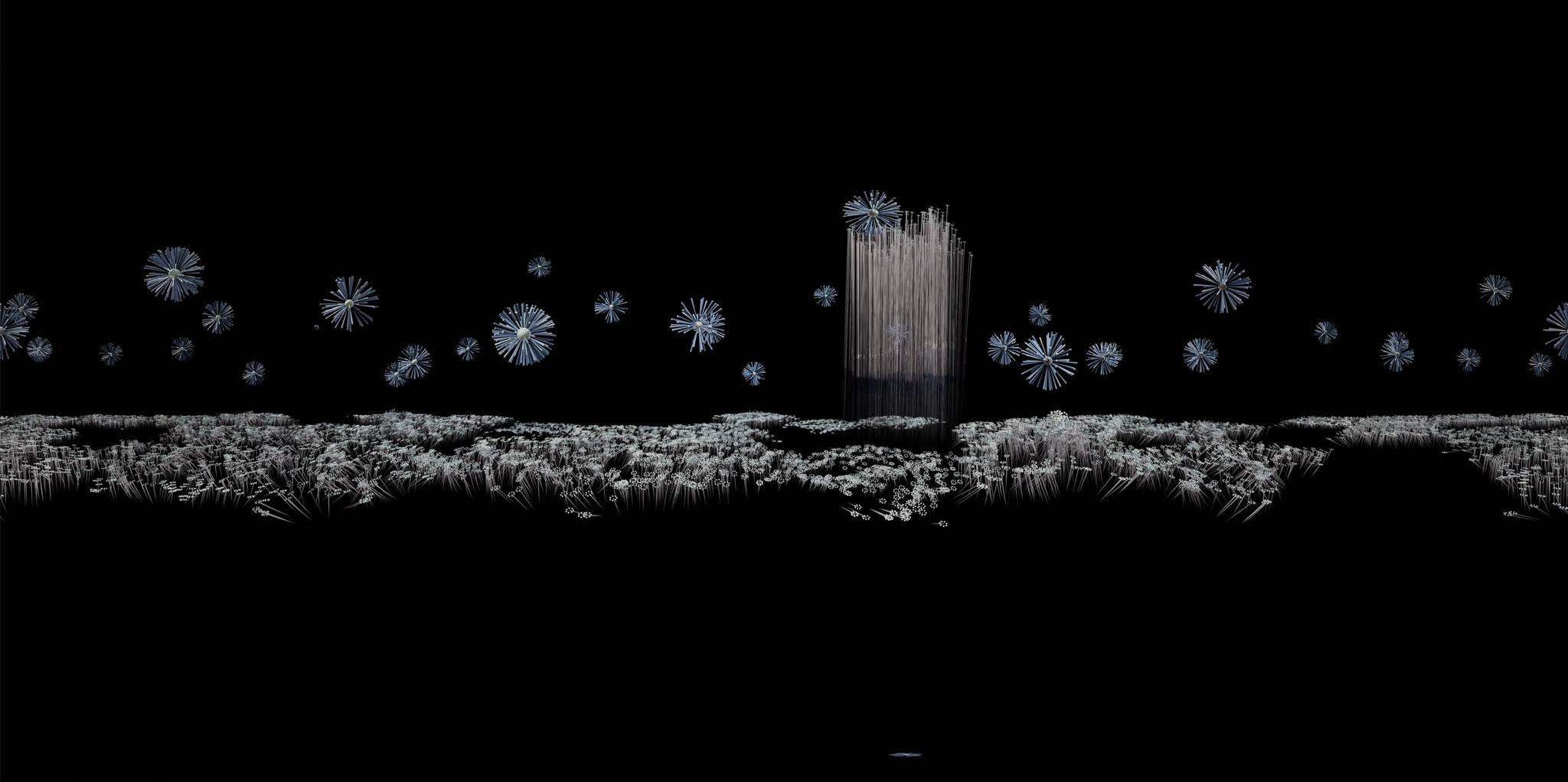}
         \caption{}
         \label{fig:five over x}
     \end{subfigure}
        \caption{Three virtual environments were used during the VR experiment: a) \textsc{Video} -- the original 360-degree recording of the live performance; b) \textsc{Mixed} -- the condition combining data visualization and original performance; c) \textsc{Symbolic} -- the visualization based on the audience data.}
        \label{fig3}
\end{figure}

We integrated recorded 360-degree video, audio, and visualized physiological data from the concert into the VR environment. To study the level of abstraction for recreation in VR, we presented VR users with the same piano performance across three conditions with abstraction levels ranging from the original video recording to abstract visualization of audience responses. The three VR conditions are as follows:

\textbf{Video}: A recorded 360-degree video of the concert performance (see Figure~\ref{fig:five over x}). Users observe the pianist's expressions and movements while viewing surrounding audience members and the concert hall architecture. We position 360° video as cinematic VR---empirical evidence shows no significant differences in presence or emotional responses compared to computer-simulated VR for passive viewing tasks~\cite{Brivio2021}.

\textbf{Mixed}: The video recording combined with abstract physiological visualizations (see Figure~\ref{fig3}~(b)). Spherical models derived from audience physiology float like dandelions, dynamically illustrating arousal states over time.

\textbf{Symbolic}: A fully abstract virtual environment based on audience physiological data (see Figure~\ref{fig3}~(c)). Performers and audience are not visible; they are represented through audio (music, applause) and abstract visualizations. Each flower represents an audience member's heartbeat (BVP) and changes size accordingly; flower height and grass height represent arousal (EDA).

\section{User Study:
Arousal Synchrony during virtual recreations}

We conducted an experiment to examine
the physiological responses of multiple recreations compared to the recording of a live concert performance. We recorded the concert (see~\autoref{sec:concert}) with a 360-degree video. During the experiment, we exposed participants to the original video recording, the mixed condition, and the fully reconstructed symbolic condition within the VR context. We conducted data collection using sensing devices (EDA and BVP sensors, consistent with those used during the concert recording) and questionnaires. The physiological data acquisition devices comprise electrodermal activity sensors and heart activity sensors.

\subsection{Study Design}

We conducted a two-phase investigation to examine whether VR concert recreations can evoke physiological arousal patterns comparable to those observed in live audiences. Phase 1 consisted of a field study at a classical music concert, during which we recorded electrodermal activity (EDA) from 50 attendees; after artifact rejection and signal quality screening, 40 participants were retained for analysis. These data served as both a behavioral ground truth and the source for aggregated arousal signals used in Phase 2. The second phase brought 22 new participants into the laboratory, where they experienced VR recreations of concert excerpts under three visualization conditions. IRB approval from the ethics committee at Keio University Graduate School of Media Design and informed consent were obtained.

Participants in Phase 2 completed all three conditions in a within-subjects repeated-measures design. The \textit{Video} condition presented realistic 360-degree footage of the concert hall and performance. The \textit{Mixed} condition overlaid real-time physiological visualizations onto this video feed, rendering aggregated audience arousal as ambient visual effects. The \textit{Symbolic} condition replaced the realistic imagery entirely with abstract, procedurally generated visuals driven by the same physiological data. To mitigate order effects, we counterbalanced condition sequences using complete counterbalancing across all six orderings based on enrollment order.

The central analytical strategy involved cross-temporal synchrony comparisons. Rather than comparing VR participants to one another, we examined whether each individual's moment-to-moment EDA aligned with the aggregated arousal trace from the original live audience. This approach allowed us to ask a more ecologically meaningful question: do VR viewers respond to the same musical passages that moved in-person attendees? By treating the Phase 1 aggregate as a fixed reference signal, we could assess whether different visualization strategies modulate not just overall arousal but its temporal correspondence with authentic audience responses.

\subsection{Participants and Apparatus}

22 participants (6 female, 14 male, 2 unknown) aged 24-34 years ($M = 27.43$, $SD = 2.62$) with 90.9\% having concert experience participated for \$10 compensation. Post-hoc power analysis confirmed adequate power (.85) for detecting medium effects ($f = 0.25$) with within-subjects design.

Equipment included HTC Vive Pro Eye HMD (90Hz, 1440×1600 per eye, 110° FOV) with 3D spatial audio and Shimmer3 GSR+ sensors for EDA/BVP recording, identical to those used at the concert.

\subsection{Procedure and Measurements}
Three counterbalanced VR conditions were tested per participant. SAM questionnaires~\cite{bradley1994measuring} assessed subjective responses before/after each condition. Physiological data (EDA: 10Hz, BVP: 200Hz) were recorded throughout 40-50 minute sessions using identical sensors from the concert. Participants were naive to visualization content.
Protocol steps: (1) informed consent and demographics collection; (2) sensor attachment to non-dominant hand with movement restrictions; (3) pre-condition SAM questionnaire; (4) 5:11 VR experience with head tracking only; (5) post-condition SAM and SUS presence questionnaires; (6) repetition for remaining conditions; (7) semi-structured interview covering overall experience, condition preferences, visualization insights, and design element evaluations.

\subsection{Statistical Analysis Framework}

Statistical analyses used Python 3.11 with SciPy/Statsmodels ($\alpha = .05$). The within-subjects design ($N = 22$) employed counterbalanced conditions with adequate power (.85) for medium effects. Due to DTW normality violations, we applied Aligned Rank Transform~\cite{wobbrock2011aligned} before ANOVAs. Effect sizes used partial eta-squared for ANOVAs and Pearson's $r$ for correlations. Multiple comparisons employed Tukey's HSD with Holm-Bonferroni correction. Primary analyses examined: (1) between-condition DTW differences, (2) live-VR physiological correlations, and (3) subjective experience measures.

\section{Results}
Having developed a system for transforming live concert physiological data into VR visualizations, we now examine whether this approach achieves cross-temporal physiological alignment between live and VR audiences. Our analysis focuses on Dynamic Time Warping correlations between live concert EDA patterns and VR participant responses, supplemented by questionnaire data and interviews to understand the subjective experience. Abstract physiological visualizations unexpectedly achieve superior synchrony compared to realistic recreations.

This section presents the pre-processing and results of analyzing participant responses through physiological data (Electrodermal Activity), questionnaire scales, and semi-structured interviews.

\vspace{-0.2cm}
\subsection{Key Findings Summary}

Our analysis reveals two key findings that challenge assumptions about presence and realism in VR design:

\textbf{1) Abstract visualizations showed higher physiological synchrony:} The \textsc{Symbolic} condition achieved the strongest correlation with live concert tonic EDA patterns ($r(258) = .96$, $p < .001$), compared to both \textsc{Mixed} ($r(258) = .92$, $p < .001$) and \textsc{Video} conditions ($r(258) = .91$, $p < .001$). Conditions with reduced visual realism while preserving physiological patterns showed higher autonomic alignment than those with greater realism.

\textbf{2) Peak synchrony occurs during musical climax periods:} All three VR conditions showed strongest physiological alignment with the live audience during the performance climax (100--160s), with the \textsc{Symbolic} condition maintaining significant correlation ($r(58) = .84$, $p < .001$) while the \textsc{Video} condition showed no significant correlation during this critical period ($r(58) = -.13$, $p = .323$).

\vspace{-0.3cm}
\subsection{Analysis of Physiological Data}
Excluding instances with unrecorded data due to technical malfunctions, we collected 47 EDA and 41 BVP recordings from 50 audience members (female: 25, male: 20, unknown: 5) at the on-site live concert, as well as 22 EDA and 22 BVP recordings from the VR experiment. By ruling out incomplete or noisy data records, we had 40 EDA and 36 BVP recordings from the recruited concert audience, resulting in the following number of data sets for each condition of the VR experiment: \textsc{Video} - 21 EDA and 21 BVP; \textsc{Mixed} - 20 EDA and 21 BVP; and \textsc{Symbolic} - 21 EDA and 21 BVP.

BVP data does not capture rapid physiological changes well due to lower sensitivity and signal artifacts. Its lower temporal resolution makes it less effective for analyzing synchrony in brief settings. We focused on EDA data for this study.

\vspace{-0.2cm}
\subsubsection{Pre-processing of Electrodermal Activity (EDA)} 

The raw EDA data for each participant was processed using a second-order Butterworth low-pass filter (0.5 Hz)~\cite{gustafsson1996determining} from the scipy.signal package to reduce electrical noise. 
Seven samples from the VR experiment and 11 samples from the concert were processed to identify and correct outliers based on quantile-derived boundaries at the 25th and 75th percentiles. Outliers were replaced using linear interpolation between the nearest valid data points surrounding each outlier group, ensuring smooth transitions in the signal. After interpolating, a moving average was applied to further smooth the data. Finally, to compensate for individual differences in the data, each EDA sample was normalized using a min-max range normalization technique. Neurokit2~\cite{makowski2021neurokit2} was used for further feature extraction.

EDA signals were decomposed into tonic (slowly changing baseline) and phasic (rapidly changing) components~\cite{boucsein2012electrodermal,cowley2016psychophysiology,dawson2000handbook}. We focused our synchrony analysis on tonic EDA, which reflects overall arousal state and is more suitable for cross-temporal comparison than the highly variable phasic component. We examined participants' physiological synchrony during the piano performance to assess the influence of three VR conditions on arousal alignment with the original live audience.

\subsubsection{Measurement of Physiological Synchrony}
We quantified physiological synchrony using Pearson Correlation (PC)~\cite{freedman2007statistics} and Dynamic Time Warping (DTW)~\cite{giorgino2009computing,salvador2007toward, tormene2009matching} at both inter-individual and group levels.

Pearson Correlation (PC) measures how two continuous signals co-vary over time and represents the linear relationship numerically, ranging from -1 (negative correlation) to 0 (no correlation) to 1 (perfect correlation). It is easy to understand, and has been widely applied in measuring synchrony~\cite{pijeira2016investigating}. The higher the PC coefficient, the greater the synchrony between the two physiological signals.

Dynamic Time Warping (DTW) is an algorithm that seeks the optimal alignment and similarity between two time series that may differ in time and speed. We employed the DTW Python library to calculate the DTW values between various physiological signals. The result provided by DTW is the distance between the two physiological signals. Therefore, the higher the DTW value between two signals, the lower their synchrony, and vice versa.
\begin{figure*}[ht]
    \centering
    \includegraphics[width=1\linewidth]{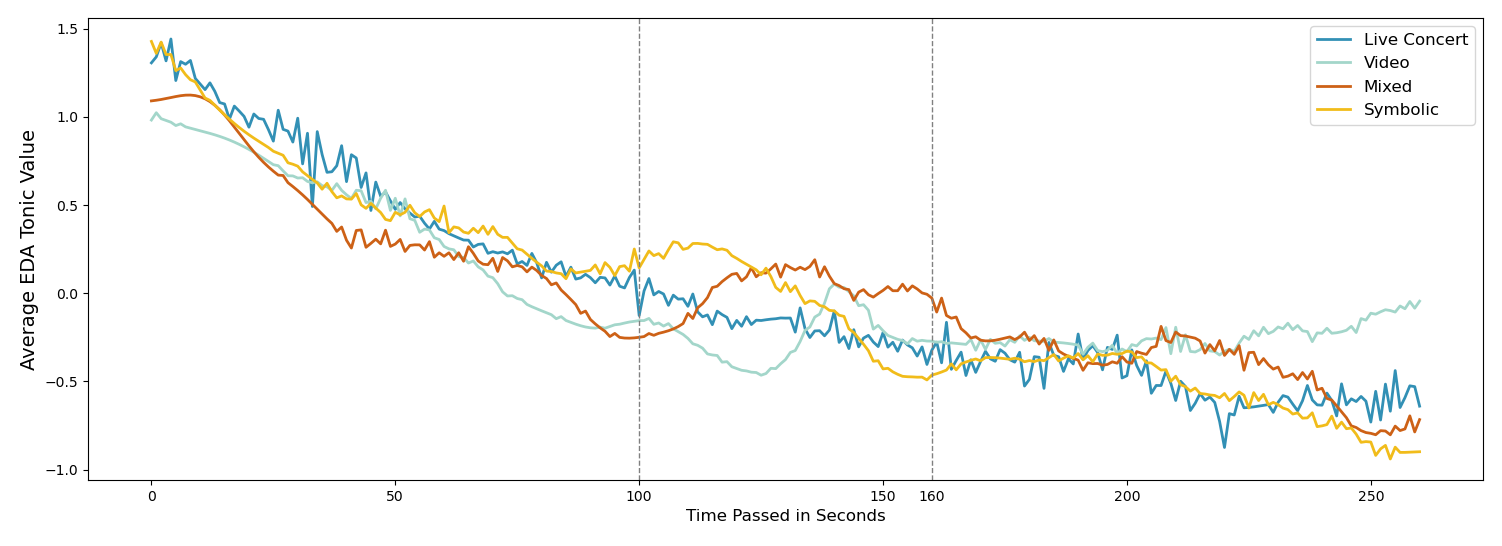}
    \caption{Comparison of average tonic EDA values over time for live concert and the three VR conditions: \textsc{Video}, \textsc{Mixed}, and \textsc{Symbolic}. The x-axis denotes time in seconds, segmented by time marks at 100 seconds and 160 seconds, dividing the experience into three phases. The time series shows an initial high response across all conditions, followed by a general decline and varied fluctuations between the 100- and 160-second marks. Significant Pearson's correlations were observed between the live concert and each VR condition: \textsc{Video}, $r(258) = .91$, $p < .001$; \textsc{Mixed}, $r(258) = .92$, $p < .001$; and \textsc{Symbolic}, $r(258) = .96$, $p < .001$, with further details shown in Table~\ref{tab:tonic}.}
    \label{fig:average tonic}
    \Description{}
\end{figure*}

\vspace{-0.2cm}

\subsection{Results of Physiological synchrony}

\subsubsection{Synchrony between Live Concert and VR Conditions (groups)}
We calculated average tonic EDA values (see Figure~\ref{fig:average tonic}) for both study phases: the live concert baseline (N=40 field study participants) and the three VR conditions (\textsc{Video}, \textsc{Mixed}, \textsc{Symbolic}) from the laboratory study (N=22 participants). Using Pearson Correlation and DTW, we analyzed synchrony between the live concert and VR conditions. Previous work suggests physiological synchrony peaks during musical climaxes~\cite{czepiel2021synchrony}. We examined correlations across different sections of the performance, with particular focus on the climax (100--160s).

Table~\ref{tab:tonic} shows Pearson correlation and DTW analyses across three time segments and the entire duration. Strong positive correlations were observed between live concert tonic EDA and all three VR conditions for the entire duration, with \textsc{Symbolic} achieving the highest correlation ($r = .96$, $p < .001$). \textsc{Symbolic} exhibited the strongest synchronization across all time segments, while \textsc{Video} was the only condition showing no significant correlation during the 100--160s climax segment ($r = -.13$, $p = .323$).

\begin{table*}[ht]
\centering
\caption{Pearson correlation and DTW Values for average EDA tonic between the live concert and each of the three VR conditions across three time segments (0--100s, 100--160s, 160--260s) and over the entire duration.}
\label{tab:tonic}
\resizebox{1\linewidth}{!}{%
\begin{tabular}{|c|ccccccccc|}
\hline
\multirow{3}{*}{\textbf{SceneTime slot}} & \multicolumn{3}{c|}{\textbf{0--100s}} & \multicolumn{3}{c|}{\textbf{100--160s}} & \multicolumn{3}{c|}{\textbf{160--260s}} \\ \cline{2-10} 
 & \multicolumn{2}{c|}{Pearson correlation} & \multicolumn{1}{c|}{\multirow{2}{*}{DTW}} & \multicolumn{2}{c|}{Pearson correlation} & \multicolumn{1}{c|}{\multirow{2}{*}{DTW}} & \multicolumn{2}{c|}{Pearson correlation} & \multirow{2}{*}{DTW} \\ \cline{2-3} \cline{5-6} \cline{8-9}
 & \multicolumn{1}{c|}{r value} & \multicolumn{1}{c|}{p value} & \multicolumn{1}{c|}{} & \multicolumn{1}{c|}{r value} & \multicolumn{1}{c|}{p value} & \multicolumn{1}{c|}{} & \multicolumn{1}{c|}{r value} & \multicolumn{1}{c|}{p value} &  \\ \hline
Video & \multicolumn{1}{c|}{0.963} & \multicolumn{1}{c|}{\textless{}.001} & \multicolumn{1}{c|}{8.317} & \multicolumn{1}{c|}{-0.129} & \multicolumn{1}{c|}{0.323} & \multicolumn{1}{c|}{6.790} & \multicolumn{1}{c|}{-0.370} & \multicolumn{1}{c|}{\textless{}.001} & 24.232 \\ \hline
Mixed & \multicolumn{1}{c|}{0.960} & \multicolumn{1}{c|}{\textless{}.001} & \multicolumn{1}{c|}{6.829} & \multicolumn{1}{c|}{-0.512} & \multicolumn{1}{c|}{\textless{}.001} & \multicolumn{1}{c|}{11.664} & \multicolumn{1}{c|}{0.452} & \multicolumn{1}{c|}{\textless{}.001} & 7.656 \\ \hline
Symbolic & \multicolumn{1}{c|}{0.969} & \multicolumn{1}{c|}{\textless{}.001} & \multicolumn{1}{c|}{4.326} & \multicolumn{1}{c|}{0.839} & \multicolumn{1}{c|}{\textless{}.001} & \multicolumn{1}{c|}{6.267} & \multicolumn{1}{c|}{0.691} & \multicolumn{1}{c|}{\textless{}.001} & 8.118 \\ \hline
\multirow{2}{*}{\textbf{SceneTime slot}} & \multicolumn{9}{c|}{\textbf{0--260s}} \\ \cline{2-10} 
 & \multicolumn{3}{c|}{r value} & \multicolumn{3}{c|}{p value} & \multicolumn{3}{c|}{DTW} \\ \hline
Video & \multicolumn{3}{c|}{0.908} & \multicolumn{3}{c|}{\textless{}.001} & \multicolumn{3}{c|}{36.036} \\ \hline
Mixed & \multicolumn{3}{c|}{0.919} & \multicolumn{3}{c|}{\textless{}.001} & \multicolumn{3}{c|}{16.545} \\ \hline
Symbolic & \multicolumn{3}{c|}{0.960} & \multicolumn{3}{c|}{\textless{}.001} & \multicolumn{3}{c|}{14.610} \\ \hline
\end{tabular}%
}
\end{table*}

\vspace{-0.1cm}
\subsubsection{Synchrony between Live Concert and VR conditions (individuals)}

\begin{figure*}[ht]
    \centering
    \includegraphics[width=1\linewidth]{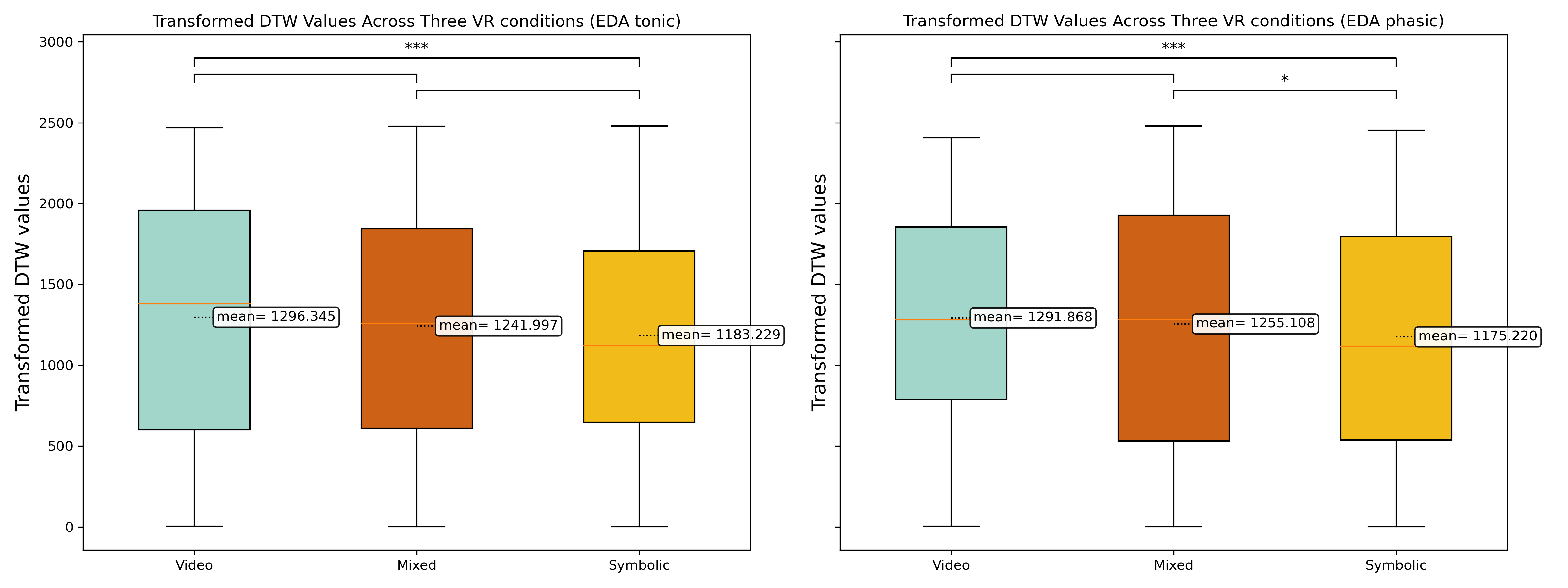}
    \caption{Distribution of transformed DTW values for tonic EDA comparisons between individual participants of the live concert and those of the three VR conditions. Each box plot shows the distribution within each group, with the mean indicated by a horizontal line. Statistical significance between groups is denoted by asterisks (*$p < .05$, **$p < .005$, ***$p < .001$). The transformed DTW values in \textsc{Symbolic} are significantly lower than in \textsc{Video} ($p=.001$), indicating higher synchrony between \textsc{Symbolic} and live concert.}
    \label{fig:DTW_boxplot}
\end{figure*}

\begin{figure*}[ht]
  \centering
  \includegraphics[height=0.28\textheight, keepaspectratio]{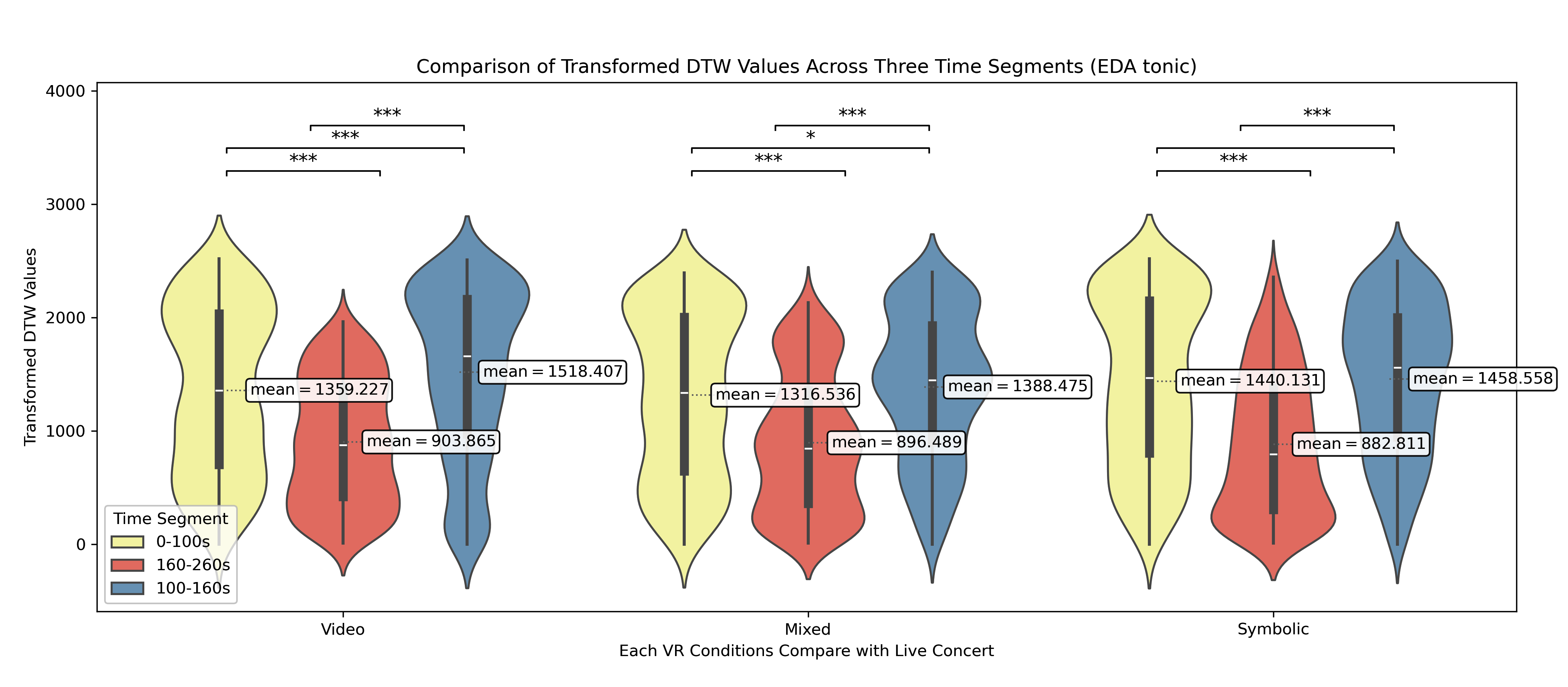}
  \caption{Comparison of transformed DTW values for tonic EDA across three VR conditions and time segments. Lower values indicate higher synchrony with live concert audience.}
  \label{fig:DTW_timesegements}
\end{figure*}

We performed DTW analysis on tonic EDA values from individuals at the live concert (40 recordings) and VR participants (\textsc{Video}: 21, \textsc{Mixed}: 20, \textsc{Symbolic}: 21). We formed pairs for DTW analysis by matching each VR participant with live concert audience members, creating 840 pairs for \textsc{Video}, 800 for \textsc{Mixed}, and 840 for \textsc{Symbolic}. A Shapiro-Wilk test found some conditions violated normality assumptions, so we applied an Aligned Rank Transform (ART)~\cite{wobbrock2011aligned} before conducting ANOVA. We examined DTW differences across VR conditions (Figure~\ref{fig:DTW_boxplot}) and time segments (Figure~\ref{fig:DTW_timesegements}).

Tonic EDA synchrony analysis using ART ANOVA revealed a significant main effect of VR condition on DTW values, $F(2, 2477) = 5.13$, $p = .006$, $\eta^2_p = .004$, 95\% CI [.001, .012]. Post-hoc Tukey HSD tests with False Discovery Rate correction showed significant differences between \textsc{Video} condition ($M = 1296.35$, $SD = 24.97$) and \textsc{Symbolic} condition ($M = 1183.23$, $SD = 24.97$), with a mean difference of 113.12, $p = .001$, Cohen's $d = 0.94$, indicating a large effect size for higher synchrony in the symbolic visualization.

Time segment analysis for tonic EDA synchrony revealed significant temporal effects across all VR conditions. The \textsc{Video} condition showed $F(2, 2517) = 126.43$, $p < .001$, $\eta^2_p = .091$, 95\% CI [.076, .108]. The \textsc{Mixed} condition demonstrated $F(2, 2397) = 118.76$, $p < .001$, $\eta^2_p = .090$, 95\% CI [.075, .107]. The \textsc{Symbolic} condition exhibited $F(2, 2517) = 108.22$, $p < .001$, $\eta^2_p = .079$, 95\% CI [.064, .096]. During the middle segment (100--160s), all conditions showed significantly higher synchronization compared to other time periods, with all pairwise comparisons surviving Bonferroni correction for multiple testing.

\subsection{Audience Questionnaires}
To assess their subjective experiences, participants rated three dimensional subjective scales using 9-point Self-Assessment Manikins (SAM) before and after each VR scene. Valence measured how positive or negative the experience was, Arousal assessed intensity, and Dominance evaluated perceived control. Repeated-measures ANOVA analysis of change scores revealed no significant differences across the three VR conditions. For Valence, $F(2, 42) = 0.03$, $p = .974$, $\eta^2_p = .001$, 95\% CI [.000, .018]. For Arousal, $F(2, 42) = 1.61$, $p = .212$, $\eta^2_p = .071$, 95\% CI [.000, .221]. For Dominance, $F(2, 42) = 0.72$, $p = .495$, $\eta^2_p = .033$, 95\% CI [.000, .142]. The absence of significant differences in self-reported emotional dimensions suggests that higher physiological synchrony was observed without conscious awareness of the visualization manipulation.

Presence assessment using the Slater-Usoh-Steed questionnaire showed mean scores of 1.95 for \textsc{Video}, 1.50 for \textsc{Mixed}, and 1.41 for \textsc{Symbolic} conditions. One-way ANOVA revealed no significant differences in presence scores across conditions, $F(2, 63) = 0.84$, $p = .437$, $\eta^2_p = .026$, 95\% CI [.000, .118]. Results indicate that the physiological synchrony effects were not mediated by differences in subjective presence, suggesting that the symbolic visualization was associated with higher physiological coupling not reflected in self-reported presence.
\subsection{Audience Interviews}
Interview transcripts were analyzed inductively by two researchers, coding 236 data points into 20 themes. Three patterns emerged across conditions.

\textbf{Condition Preferences:} The \textsc{Symbolic} condition was preferred by half of participants (11/22), who described feeling calm, immersed, and better able to focus on the music. The \textsc{Video} condition was valued for realism (8/22 preferred), though 7 participants found it boring or overly conventional. The \textsc{Mixed} condition appealed to 6 participants for its novelty, but 12/22 found the combination overwhelming or unnatural.

\textbf{Calmness and Engagement:} Over half of participants (13/22) spontaneously mentioned ``calm'' or ``calming'' when describing their experiences, particularly in the \textsc{Symbolic} condition. One participant explained: \textit{``It gives me a feeling of resonance among the image, music, and performance...I have arrived in the world of music.''} (P14)

\textbf{Visualization Perception:} Six participants recognized heartbeat-like rhythms in the 3D object pulsations, and five associated the visualizations with audience presence: \textit{``I sense like maybe it's the spirit of the people, how they are responding, their presence...it's beating. It's like a heart beating.''} (P08)

\section{Discussion}
In this section, we interpret our findings by first examining how musical highlights structure moments of heightened physiological synchrony. We then reflect on what these results imply for HCI evaluation, before turning to the unexpected advantage of abstract visualizations in aligning arousal with a live audience. Finally, we argue that physiological synchrony offers a valuable complementary lens for evaluating immersive experiences beyond traditional presence measures.

\subsection{Highlights in Performances Result in Higher Synchrony}
We segmented time based on the structure of classical music and feedback from the performer. The 100--160s interval represents the climax of the music. Our results, as shown in Figure~\ref{fig:DTW_timesegements}, indicate that DTW values for both phasic EDA and tonic EDA are significantly lower in the 100--160s segment compared to other time segments (0--100s and 160--260s). Data suggest that both tonic EDA and phasic EDA datasets achieve higher synchrony during the climax section across all VR conditions (\textsc{Video}, \textsc{Mixed}, and \textsc{Symbolic}).

To investigate the underlying reasons for higher synchrony during musical climaxes: is it the expressive quality of the music, its volume, character, or tempo that most effectively produces physiological activation? This question will guide our future research directions. We categorized physiological data into three time-based segments, reflecting the physiological progression during the music. Going forward, we can further dissect and analyze the data at specific time points, considering musical characteristics or transient events.

\subsection{Methodological Implications for HCI Evaluation}

Our approach demonstrates how physiological synchrony analysis can extend existing HCI evaluation frameworks beyond traditional presence measures and self-report questionnaires. By measuring cross-temporal autonomic alignment between live and virtual audiences, we provide a non-disruptive methodology that captures moment-to-moment physiological responses without interrupting the experience. This addresses a fundamental limitation in VR cultural recreation evaluation, where traditional measures fail to capture the complex social dynamics that make live events compelling.

Abstract visualizations achieving superior physiological synchrony challenges core assumptions in presence research about the relationship between visual fidelity and experience quality. Results suggest that HCI evaluation methodology should distinguish between cognitive measures of realism and embodied measures of physiological engagement. Our Dynamic Time Warping approach enables researchers to evaluate VR experiences based on their ability to recreate specific physiological patterns rather than subjective impressions of presence, providing a more objective and specific measure of recreation effectiveness.

This methodology has broader implications for evaluating biometric-responsive systems, social VR experiences, and any application where shared physiological states are relevant to experience quality. The approach can be adapted to different physiological measures (heart rate variability, respiratory patterns) and cultural contexts, offering a generalizable framework for assessing autonomic alignment in immersive experiences.

\subsection{Abstraction Boosts Arousal Synchrony with Live Audience}
In our synchrony analysis, we computed group- and individual-level alignment between VR conditions (\textsc{Video}, \textsc{Mixed}, \textsc{Symbolic}) and the live concert using tonic EDA with PC and DTW.
Across metrics, \textsc{Symbolic} produced higher synchrony than \textsc{Mixed} and \textsc{Video}.
This aligns with somaesthetic appreciation design~\cite{hook2016somaesthetic}: abstract, body-centered representations focus attention on felt physiological patterns rather than realistic detail, yielding stronger embodied engagement.
Participants described \textsc{Video} as "real" and "being there," yet also reported boredom; \textsc{Symbolic} 3D elements were associated with novelty and engagement that standard presence scales miss.
From 0–260s the average tonic EDA trends are similar to the concert across all VR conditions; \textsc{Symbolic} shows the lowest DTW and strongest Pearson correlation to live, and individual analyses likewise favor \textsc{Symbolic} (see Figure~\ref{fig:DTW_timesegements}).

\vspace{-0.3cm}
\subsection{Why Abstract Visualizations Showed Higher Synchrony}

Abstract visualizations produced higher synchrony than realistic video---a counterintuitive result with several possible explanations. First, abstract representations avoid uncanny valley effects; participants engage with them as ambient information displays rather than comparing them to expectations of real footage. Second, the Symbolic condition reduces perceptual load, freeing attention for the music and rhythmic visualization patterns. Third, novelty may have captured sustained attention, though longitudinal studies should test whether this advantage persists.

We cannot rule out that our visualizations were simply more aesthetically engaging independent of their physiological basis. Future work should compare physiological visualizations to matched non-physiological animations to isolate this factor.

For designers, these findings suggest that realistic recreation is not always optimal. Abstract physiological representations offer a promising direction for collective engagement in asynchronous settings. Cross-temporal physiological synchrony provides a non-intrusive evaluation methodology complementing questionnaire measures. More concretely, designers should consider abstract physiological visualizations as a viable alternative to realistic recreation when the goal is to foster collective engagement in asynchronous VR experiences. Rather than viewing abstraction as a limitation of fidelity, this work suggests it can be a deliberate design choice that reduces perceptual load and avoids uncanny valley effects. VR practitioners can evaluate such recreations using cross-temporal physiological synchrony metrics, offering an unobtrusive complement to presence questionnaires. The optimal abstraction level likely varies by domain---a framework worth systematic exploration across different genres and cultural contexts.

\subsection{Physiological Synchrony as Evaluation Method}
Traditional measures (presence, SAM) showed no significant differences across conditions, despite clear physiological differences. This supports performance research advocating for evaluation methods that capture embodied, shared experiences beyond presence measures~\cite{benford2013performance}. Physiological synchrony enables observation of subtle arousal responses that traditional methods miss, extending typical face-to-face applications of synchrony measurement to cross-temporal VR contexts.

While our findings emerge from a specific musical context, the underlying mechanism (leveraging abstract physiological representations to bridge temporal gaps between shared experiences) may extend to other domains where collective arousal plays a central role. Theater performances, live dance, and sporting events all generate measurable audience synchrony that could, in principle, be captured and re-presented through similar visualization approaches, though the optimal abstraction level likely varies with the nature of the experience and its emotional dynamics. Collaborative VR environments present a particularly intriguing direction, where real-time physiological coupling might enhance remote co-presence in ways that warrant systematic investigation. We are cautious about overgeneralizing from a single concert study, however; future work should explore whether the advantages we observed for symbolic visualization persist across these diverse contexts or reflect something particular to musical immersion.

\section{Limitations}
Comparing live concert physiological data with VR data introduces variance in participants, timing, location, and medium. Standard laboratory validation using public stimulus databases would exclude the complex audience reactions inherent to live events. Questionnaires collected only before and after each experience limit our ability to validate physiological patterns against moment-to-moment subjective states.

Our study did not collect resting-state physiological baselines prior to each condition, which limits our ability to account for individual differences in tonic arousal levels. While our focus on synchrony patterns partially mitigates this concern, since we examine temporal covariation rather than absolute magnitude, future work should incorporate baseline periods. Additionally, the three VR conditions differed along multiple perceptual dimensions beyond visual abstraction, including presence of human figures, visual complexity, and spatial density. We cannot fully disentangle which specific features drove the observed differences, though our design reflects the practical reality that abstraction level covaries with these properties in most artistic visualization approaches.

\section{Conclusion}
We present cross-temporal physiological synchrony analysis as a methodology for evaluating VR cultural recreations by measuring autonomic alignment between live and virtual audiences. Using Pearson correlation and DTW, we found that VR users showed higher synchrony with live concert audience members when exposed to abstract physiological visualizations compared to realistic video.

The \textsc{Symbolic} condition achieved the highest arousal synchrony ($r = .96$) with the live audience, both at group and individual levels. Peak synchrony occurred during the musical climax (100--160s). These findings challenge assumptions about presence and realism in VR design, suggesting that abstract visualizations may better preserve the physiological dynamics of collective cultural experiences.

\begin{acks}

Generative AI tools were used at the paragraph level to improve readability and style of text originally envisioned and written by the authors, and to adjust figures (e.g., improving clarity, layout, or labeling) based on author-created originals. The AI assistance was limited to copy-editing and visual refinement; all conceptual content, study design, analysis, and interpretation were created by the authors. All AI-assisted text and image adjustments were carefully reviewed, revised, and approved by the human authors, who remain fully responsible for the final manuscript. No confidential or proprietary data were provided to the AI system.
\end{acks}


\bibliographystyle{ACM-Reference-Format}
\bibliography{sample-base}

\end{document}